# Double C-NOT attack on a single-state semi-quantum key distribution protocol and its improvement

**Jun Gu [1], Tzonelih Hwang[*]**

*Department of Computer Science and Information Engineering, National Cheng Kung University, No. 1, University Rd., Tainan City, 70101, Taiwan, R.O.C.*

[1] isgujun@163.com

[*]**Responsible for correspondence:**

Tzonelih Hwang

Distinguished Professor

Department of Computer Science and Information Engineering,

National Cheng Kung University,

No. 1, University Rd.,

Tainan City, 70101, Taiwan, R.O.C.

Email: hwangtl@ismail.csie.ncku.edu.tw

TEL: +886-6-2757575 ext. 62524

**Abstract**

Recently, Zhang et al. proposed a single-state semi-quantum key distribution protocol (Int. J. Quantum Inf, 18, 4, 2020) to help a quantum participant to share a secret key with a classical participant. However, this study shows that an eavesdropper can use a double C-NOT attack to obtain parts of the final shared key without being detected by the participants. To avoid this problem, a modification is proposed here.

**Keywords:** Quantum key distribution; Semi-quantum; Double C-NOT attack

## 1. Introduction

Quantum key distribution (QKD) protocol [1] is designed to help the participants share a secret key without being eavesdropped. However, in most of the existing QKD protocols [2-4], all the participants need to have lots of quantum capabilities, such as quantum register, quantum joint measurement, and so on. To help the participants who just have restricted quantum capacities can be involved in the QKD, the semi-quantum key distribution (SQKD) protocol is proposed.

In 2007, an SQKD protocol was proposed by Boyer et al. [5]. With their protocol, a classical participant with restricted quantum capacities can share a secret key with a quantum participant who has unrestricted quantum capacities. According to Boyer et al.'s definition, the classical participants are restricted to perform parts of the following quantum operations: (1) generate qubits in Z-basis $\{|0\rangle, |1\rangle\}$, (2) measure qubits with the Z-basis, (3) reorder the qubits via different quantum delay lines, and (4) send or reflect the qubits. Subsequently, these definitions have been widely used in the following proposed semi-quantum protocols [6-12].

Recently, Zhang et al. [13] proposed a single-state semi-quantum key distribution protocol. They claimed that the proposed SQKD protocol can ensure a quantum participant and a classical participant share a secret key without being eavesdropped. However, this study shows that Zhang et al.'s SQKD protocol suffers from a double C-NOT attack which can reveal parts of the final shared key to an eavesdropper. To solve this problem, a modification is proposed.

The rest of this paper is organized as follows. Section 2 briefly reviews Zhang et al.'s SQKD protocol. Section 3 shows that an eavesdropper can obtain parts of the final shared key by a double C-NOT attack and then proposes an improvement to avoid it. At last, a conclusion is given in Section 4.

## 2. A brief review of Zhang et al.'s SQKD

In Zhang et al.'s protocol [13], four kinds of single photons $\{|0\rangle, |1\rangle, |+\rangle = \frac{1}{\sqrt{2}}(|0\rangle+|1\rangle), |-\rangle = \frac{1}{\sqrt{2}}(|0\rangle-|1\rangle)\}$ are used. Assume that Bob is a classical participant who just has the above quantum capacities {(1), (2), (4)} and Alice is a quantum participant with unrestricted quantum capacities. Then Zhang et al.'s SQKD protocol can be described as follows:

**Step 1**: Alice generates $n$ single photons in $|+\rangle$ and sends these single photons to Bob one by one.

**Step 2**: Bob generates a random bit sequence $K_B = \{k_{B1}, k_{B2}, \cdots, k_{Bn}\}$. For the $i$ th ($1 \leq i \leq n$) particle received, Bob chooses one of the two following cases according to $k_{Bi}$:

Case (a). If $k_{Bi}=0$, Bob reflects this particle to Alice directly.

Case (b). If $k_{Bi}=1$, Bob measures this particle with Z-basis and then sends a single photon in the state $|0\rangle$ back to Alice instead.

**Step 3**: For each particle sent back, Alice randomly uses Z-basis $\{|0\rangle, |1\rangle\}$ or X-basis $\{|+\rangle, |-\rangle\}$ to measure it. Then she generates a value sequence $K_A = \{k_{A1}, k_{A2}, \cdots, k_{An}\}$ according to the measurement results. That is, for the $i$th ($1 \leq i \leq n$) particle, $k_{Ai}$ is decided as follows.

(1) If the measurement result is $|1\rangle$, $k_{Ai} = 0$.

(2) If the measurement result is $|-\rangle$, $k_{Ai} = 1$.

(3) Otherwise, $k_{Ai} = -1$.

**Step 4**: Alice announces all the positions where $k_{Ai} = -1$ and discards these values in $K_A$ to obtain $K'_A$. Then Bob discards the corresponding bits in $K_B$ to derive $K'_B$.

**Step 5**: To check the eavesdropping, Bob randomly chooses half bits in $K'_B$ and announces their positions and values. Subsequently, Alice checks whether the announced positions and values are the same as $K'_A$ or not. If the error rate exceeds a predetermined value, this protocol will be aborted. Otherwise, Alice and Bob discard the announced bits in $K'_A$ and $K'_B$ to obtain the final shared key $K_{AB}$, respectively.

## 3. Double C-NOT attack and counterattack on Zhang et al.'s protocol

Zhang et al. claimed that the above SQKD protocol can ensure the final shared key is secure. However, this section points out that Zhang et al.'s SQKD protocol suffers from a double C-NOT attack. That is, an eavesdropper Eve can use a double C-NOT attack to obtain parts of the final shared key without being detected. Besides, to avoid this loophole, a simple modification is proposed here.

**3.1. The Double C-NOT Attack on Zhang et al.'s SQKD protocol**

The processes of the double C-NOT attack on Zhang et al.'s protocol can be simply described as follows. For each particle sent from Alice to Bob, Eve performs a C-NOT operation ($C-NOT = |00\rangle\langle 00| + |01\rangle\langle 01| + |10\rangle\langle 11| + |11\rangle\langle 10|$) on both the transmitted particle and a target particle generated by herself. Then, Eve sends the transmitted particle to Bob. After Bob sends a particle back in Step 2, Eve performs a C-NOT operation on this particle and the corresponding target particle again. Subsequently, Eve uses the target particle to judge whether the particle sent back from Bob is the original one or not. According to this, Eve can obtain parts of $K_{AB}$.

For example, as is shown in Table 1, assume that $q_c$ is the particle sent from Alice to Bob. Eve generates a target particle $q_t$ in the state $|0\rangle$ and performs the 1st time C-NOT operation on $\{q_c, q_t\}$ to obtain two qubits $\{q_c^1, q_t^1\}$. Eve stores $q_t^1$ and sends $q_c^1$ to Bob. Afterward, in Step 2, Bob performs the case (a) or (b) on $q_c^1$ according to $K_B$. If Bob performs the case (a) on $q_c^1$, the measurement result is named as $q_c^2$ and then Bob sends a qubit $q_c^3 = |0\rangle$ back to Alice. Otherwise, Bob directly reflects the qubit $q_c^1$ to Alice and here we also use $q_c^3$ to represent this qubit for further discussion. For $q_c^3$ sent back, Eve performs the 2nd time C-NOT operation on $\{q_c^3, q_t^1\}$ to obtain $\{q_c^4, q_t^2\}$ and sends $q_c^4$ to Alice. Subsequently, Eve measures $q_t^2$ with Z-basis. If the measurement result is $|1\rangle$, it means that the corresponding bit in $K_B$ must be '1'. Moreover, according to Table 1, we can find that no matter which operation is chosen by Bob in Step 2, $q_c^4$ always matches the case result. That is, if Bob chooses the case (a), $q_c^4$ is in $|+\rangle$ which is the expected result of the case (a) in the original protocol. Similarly, if Bob chooses the case (b), $q_c^4$ will be in $|0\rangle$ which is the same as expectation. Hence, this attack cannot be detected by the participants during the eavesdropping detection process. That means Eve can use this attack to obtain parts of $K_{AB}$ without being detected.

Table 1. States transformation during the double C-NOT attack

| Initial | 1st C-NOT | $K_B$ | Bob's operation | Particle sent back | 2nd C-NOT |
|---|---|---|---|---|---|
| $q_c, q_t$ | $q_c^1, q_t^1$ | | $q_c^2, q_t^1$ | $q_c^3, q_t^1$ | $q_c^4, q_t^1$ |
| $\|+0\rangle_{q_c q_t}$ | $\|\Phi^+\rangle_{q_c^1 q_t^1}$ | 0 | Reflect $\|\Phi^+\rangle_{q_c^2 q_t^1}$ | $\|\Phi^+\rangle_{q_c^3 q_t^1}$ | $\|+0\rangle_{q_c^4 q_t^1}$ |
| | | 1 | measure $\Rightarrow \|00\rangle_{q_c^2 q_t^1}$ | $\|00\rangle_{q_c^3 q_t^1}$ | $\|00\rangle_{q_c^4 q_t^1}$ |
| | | | measure $\Rightarrow \|11\rangle_{q_c^2 q_t^1}$ | $\|01\rangle_{q_c^3 q_t^1}$ | $\|01\rangle_{q_c^4 q_t^1}$ |

**3.2. A solution to avoid Double C-NOT attack on Zhang et al.'s SQKD protocol**

As mentioned above, because Eve can distinguish parts of the particles measured by Bob where the measurement result of $q_c^1$ is $|1\rangle$, she can obtain parts of $K_{AB}$. Hence, if Bob and Alice discard all the bits in $K_{AB}$ where the corresponding measurement result of $q_c^1$ is $|1\rangle$, then this problem can be solved. The improved protocol is then described as follows.

**Step 1'-3'** are the same as **Step 1-3** in Section 2.

**Step 4'**: Alice announces all the positions where $k_{Ai} = -1$ and Bob announces all the positions where the measurement results are $|1\rangle$ in Step 2. Subsequently, Alice and Bob discard all the bits in the above positions in $K_A$ and $K_B$ to obtain $K'_A$ and $K'_B$, respectively.

**Step 5'** is the same as **Step 5** in Section 2.

## 4. Conclusions

This paper points out a double C-NOT attack on Zhang et al.'s SQKD protocol. With this attack, an eavesdropper can obtain parts of the final shared key without being detected. To solve this problem, a modification without needing the involved classical participant to have any extra quantum capacities is proposed.

## Acknowledgment

We would like to thank the Ministry of Science and Technology of the Republic of China, Taiwan for partially supporting this research in finance under the Contract No. MOST 109-2221-E-006-168-; No. MOST 108-2221-E-006-107-.

## References

[1] Charles H. Bennet and Gilles Brassard, "Quantum cryptography: Public key distribution and coin tossing," in *Proceedings of the IEEE International Conference on Computers, Systems and Signal Processing, Bangalore, India*, 1984, pp. 175-179.

[2] Frédéric Grosshans, Gilles Van Assche, Jérôme Wenger, Rosa Brouri, Nicolas J Cerf, and Philippe Grangier, "Quantum key distribution using gaussian-modulated coherent states," *Nature,* vol. 421, no. 6920, pp. 238-241, 2003.

[3] Hoi-Kwong Lo, Xiongfeng Ma, and Kai Chen, "Decoy state quantum key distribution," *Physical review letters,* vol. 94, no. 23, p. 230504, 2005.

[4] Charles Ci Wen Lim, Christopher Portmann, Marco Tomamichel, Renato

Renner, and Nicolas Gisin, "Device-independent quantum key distribution with local Bell test," *Physical Review X,* vol. 3, no. 3, p. 031006, 2013.

[5] Michel Boyer, Dan Kenigsberg, and Tal Mor, "Quantum key distribution with classical Bob," *Physical review Letters* vol. 99, p. 140501, 2007.

[6] Michel Boyer, Ran Gelles, Dan Kenigsberg, and Tal Mor, "Semiquantum key distribution," *Physical Review A,* vol. 79, no. 3, p. 032341, 2009.

[7] Xiangfu Zou, Daowen Qiu, Lvzhou Li, Lihua Wu, and Lvjun Li, "Semiquantum-key distribution using less than four quantum states," *Physical Review A,* vol. 79, no. 5, p. 052312, 2009.

[8] Qin Li, Wh Chan, and Dong-Yang Long, "Semiquantum secret sharing using entangled states," *Physical Review A,* vol. 82, no. 2, p. 022303, 2010.

[9] Michel Boyer and Tal Mor, "Comment on "Semiquantum-key distribution using less than four quantum states"," *Physical Review A,* vol. 83, no. 4, p. 046301, 2011.

[10] Wang Jian, Zhang Sheng, Zhang Quan, and Tang Chao-Jing, "Semiquantum key distribution using entangled states," *Chinese Physics Letters,* vol. 28, no. 10, p. 100301, 2011.

[11] Jian Wang, Sheng Zhang, Quan Zhang, and Chao-Jing Tang, "Semiquantum secret sharing using two-particle entangled state," *International Journal of Quantum Information,* vol. 10, no. 05, p. 1250050, 2012.

[12] Walter O Krawec, "Mediated semiquantum key distribution," *Physical Review A,* vol. 91, no. 3, p. 032323, 2015.

[13] Wei Zhang, Daowen Qiu, and Paulo Mateus, "Single-state semi-quantum key distribution protocol and its security proof," *International Journal of Quantum Information,* vol. 18, no. 04, p. 2050013, 2020.